\documentclass[conference]{IEEEtran}
\IEEEoverridecommandlockouts
\usepackage{cite}
\usepackage{amsmath,amssymb,amsfonts}
\usepackage{algorithmic}
\usepackage{graphicx}
\usepackage{textcomp}
\usepackage{xcolor}

\usepackage{url}

\usepackage{graphicx}
\usepackage{nicefrac}

\usepackage{cite}
\usepackage{amsmath, amssymb, amsfonts}
\usepackage{algorithmic}
\usepackage{graphicx}
\usepackage{textcomp}
\usepackage{xcolor}
\usepackage{makecell} 

\usepackage{graphicx}
\usepackage{enumitem}
\usepackage{booktabs, ragged2e}
\usepackage[flushleft]{threeparttable}
\usepackage{colortbl}
\usepackage{booktabs}
\usepackage{tabularx}
\usepackage{multirow}
\usepackage{hyperref}

\def\BibTeX{{\rm B\kern-.05em{\sc i\kern-.025em b}\kern-.08em
    T\kern-.1667em\lower.7ex\hbox{E}\kern-.125emX}}
\begin{document}

\title{Single-Channel Target Speech Extraction \\ Utilizing Distance and Room
  Clues\\
}

\author{
    \IEEEauthorblockN{ Runwu Shi, Zirui Lin, Benjamin Yen, Jiang Wang, Ragib Amin Nihal, Kazuhiro Nakadai}\\[-0.8em]
    \IEEEauthorblockA{ \textit{Department of Systems and Control Engineering, Institute of Science Tokyo, Tokyo, Japan}}\\[-0.8em]
    { \{shirunwu, linzirui, wangjiang, ragib, nakadai\}@ra.sc.e.titech.ac.jp, benjamin.yen@ieee.org}
}
\maketitle

\begin{abstract}
    This paper aims to achieve single-channel target speech extraction (TSE) in enclosures utilizing distance clues and room information. Recent works have verified the feasibility of distance clues for the TSE task, which can imply the sound source's direct-to-reverberation ratio (DRR) and thus can be utilized for speech separation and TSE systems. However, such distance clue is significantly influenced by the room's acoustic characteristics, such as dimension and reverberation time, making it challenging for TSE systems that rely solely on distance clues to generalize across a variety of different rooms. To solve this, we suggest providing room environmental information (room dimensions and reverberation time) for distance-based TSE for better generalization capabilities. Especially, we propose a distance and environment-based TSE model in the time-frequency (TF) domain with learnable distance and room embedding. Results on both simulated and real collected datasets demonstrate its feasibility. Demonstration materials are available at \url{https://runwushi.github.io/distance-room-demo-page/}.
\end{abstract}


\section{Introduction}
Target speech extraction (TSE) is the task that utilizes
  speaker-related information as auxiliary clues to extract target speech from an audio mixture consisting of multiple speakers. Such speaker clues can be categorized into physiological and spatial information about the speaker. Currently, physiological clues dominate research in TSE, which includes enrolled voice
  \cite{Yang-TargetSpeakerExtractionDirectly-2024b, Yang-CoarseFineTargetSpeakerExtraction-2024,
  Yang-TargetSpeakerExtractionUltraShort-2023}, facial and lip movement \cite{Mu-SeparateSpeechChainCrossModal-2024},
  and even gestures while speaking
  \cite{Pan-SpeakerExtractionCoSpeechGestures-2022}. Another speaker clue is spatial-related
  information, mainly referring to the direction of arrival (DOA), which has been explored in the early stages of TSE. Specifically, studies use microphone arrays to enhance speech from the speaker's direction
  \cite{Flanagan-ComputersteeredMicrophoneArraysSound-1985, Zmolikova-NeuralTargetSpeechExtraction-2023},
  as well as neural TSE systems \cite{Gu-NeuralSpatialFilterTarget-2019a,
  Gu-MultiModalMultiChannelTargetSpeech-2020}. Direction clues offer several
  advantages over physiological clues. For example, direction clues do not require ideal speaker-related information, which may not always be readily available, and relying on direction clues circumvents the need to handle sensitive biometric data such as the target speaker's voice \cite{Noe-AdversarialDisentanglementSpeakerRepresentation-2021}.
  However, direction clues present challenges because they are less deterministic and are less effective in separating speakers with the same direction. 
  
  Despite these limitations, we argue that there is still room for further exploration of spatial-based TSE, particularly considering that another spatial clue: the target speaker's distance, has been underdeveloped in TSE for a considerable period. Unlike DOA-based TSE, which requires DOA estimations from a microphone array, distance-based methods can be realized in single-channel, providing more flexibility. It should be noted that distance-based methods are often designed for reverberant enclosures since distance clues rely on the transfer properties between the speaker and the microphone in enclosed spaces, where the direct-to-reverberation ratio (DRR) decreases as the target-to-microphone distance increases. To explain, the late reverberation component is less affected by distance, while the direct and early components are inversely proportional to it \cite{Kushwaha-SoundSourceDistanceEstimation-2023, Neri-SpeakerDistanceEstimationEnclosures-2024,
  Patterson-DistanceBasedSoundSeparation-2022}.
  
  Few studies have used distance information in TSE or speech separation (SS). The first attempt is Distance Based Sound Separation (DSS) \cite{Patterson-DistanceBasedSoundSeparation-2022}, in which a basic recurrent neural network (RNN) model is used to separate the single-channel mixed audio into near and far groups according to a static distance threshold such as 1.5m.
  To make such threshold-based methods more flexible,
  \cite{Gu-ReZeroRegioncustomizableSoundExtraction-2023} proposes a multichannel region based sound extraction system, in which DOA and distance clues are utilized simultaneously. Furthermore, this study also proposed a dynamic distance threshold to control the boundary between the near and far groups directly.
 \begin{figure}[!t]
    \begin{minipage}[b]{1.0\linewidth}
      \centering
      \centerline{\includegraphics[width=8.0cm]{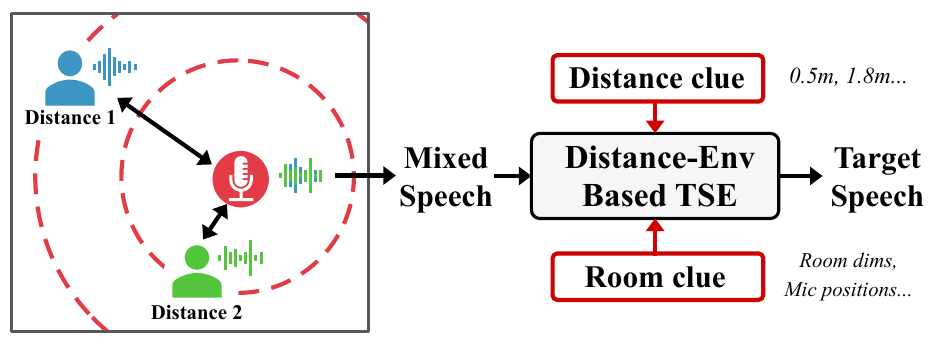}}
    \end{minipage}
    \caption{Target speech extraction using distance and room clues.}
    \label{fig:one}
    \vspace{-2em}
  \end{figure}
  Following DSS, \cite{Lin-FocusSoundYouMonaural-2023} also adopts the idea of a static distance threshold for TSE. This method contains a speaker encoder that only outputs the speaker embedding within the distance threshold, thereby the distance clue is indirectly utilized, implying this method still classifies as enrolled voice-based TSE. 
  
Unlike previous approaches, \cite{shi2024distance} introduced a distance-based TSE that relies solely on distance clues for single-channel TSE. This model outputs the speaker located near the query distance rather than separating the mixture into near and far groups. However, distance clues are highly sensitive to room acoustic parameters such as microphone placement, room dimensions, and reverberation time. Since the model does not consider these factors, its ability to generalize across different environments is significantly limited. Moreover, its performance in real-world scenarios has yet to be verified.

To address this, we explicitly incorporate room acoustic characteristics into the distance-based TSE system by designing a TF domain model that utilizes learnable distance and room embeddings. To validate this approach, we built both simulated and real-world datasets, including a location-wise real room impulse response (RIR) dataset and real-world speech recordings. The experimental results demonstrate the feasibility of our method. The remainder of this paper is organized as follows: Section 2 describes the proposed method, Section 3 presents the datasets and experimental results, and Section 4 concludes the paper.
  \section{Methodology}
  \subsection{Problem Formulation}
  Assuming $K$ speakers in an enclosed room, and denote $s_{k}(t)$, $x_{k}(t)$ and $h_{k}(t)$ as the $k$th speaker's anechoic speech, reverberant speech, and RIR, respectively. The reverberant speech $x_{k}(t)$ can be formulated as
  \begin{align}
    x_{k}(t) = s_{k}(t) \, \ast \, h_{k}(t), \label{equation:first}
  \end{align}
  where ${\ast}$ represents the convolution operator. For a single-channel microphone in the room. The collected signal mixture can be represented as
  \begin{align}
    y = \sum_{k=1}^{K}x_{k}(t). \label{equation:second}
  \end{align}
  For the proposed distance-based TSE system, it is essential to extract the target speech given a query distance and identify the presence or absence of speakers at the same time. The extraction process can be depicted as
\begin{align}
y \xrightarrow{d_q}\sum_{k}x_{k}(t), k \text{ s.t. } |d_{k} - d_{q}| \leq r_{spk}.
\end{align}
  where the $d_{q}$ is the query distance, $d_{k}$ means the relative distance between the microphone and the $k$th speaker, and $r_{spk}$ represents the speaker distance range centered with the ground truth speaker distance. Thus the boundary between the presence and the absence will be $r_{spk}$. Given a query distance $d_{q}$, the model outputs the target speech if $d_{q}$ falls within a single speaker's
  range and aggregates speech from all speakers when $d_{q}$ overlaps multiple ranges. Moreover, the model needs to output zero if the query distance $d_{q}$ falls out of the range $r_{spk}$, depicted as
  \begin{align}
    y \xrightarrow{d_q}0, \forall k, \ |d_{k} - d_{q}| > r_{spk}.
  \end{align}
  \begin{figure}[!tb]
    \begin{minipage}[b]{1.0\linewidth}
      \centering
      \centerline{\includegraphics[width=8.5cm]{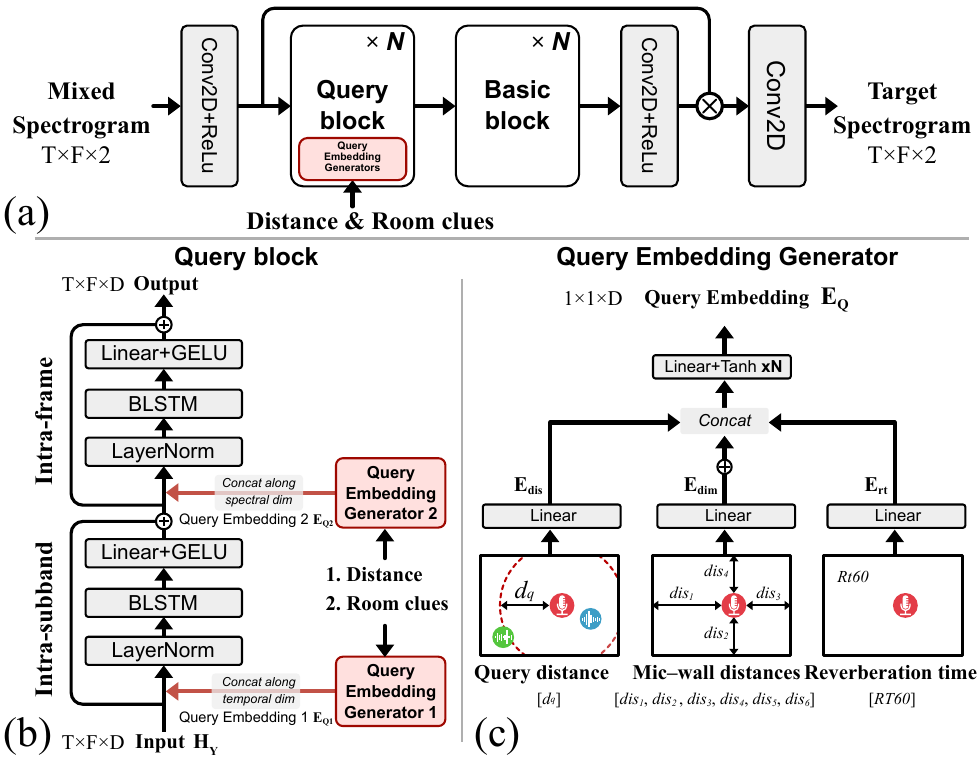}}
    \end{minipage}
    \caption{Flowchart of the proposed method showing (a) Overall structure, (b) Structure of Query block, (c) Structure of Query embedding generator.}
    \label{fig:two}
    \vspace{-2em}
  \end{figure}

  \subsection{Proposed Model}
The proposed TSE model follows the clue encoder combined with speech extraction module strategy in the TF domain \cite{Zmolikova-NeuralTargetSpeechExtraction-2023}. The overall structure is shown in Fig~\ref{fig:two} (a). The input feature of the model is the concatenated real and imaginary components $\mathbf{Y}\in \mathbb{R}^{2 \times T \times F}$ obtained through short-time Fourier transform (STFT), which is then processed by the encoder consisting of the initial 2D convolution layer with $3 \times 3$ kernel and the global layer normalization followed by the rectified linear unit (ReLu) activation function to obtain the non-negative $D$-dimensional TF embeddings $\mathbf{H_Y}\in \mathbb{R}^{D \times T \times F}$ \cite{Wang-TFGridNetIntegratingFullSubBand-2023, Yang-CoarseFineTargetSpeakerExtraction-2024}. The TF embeddings $\mathbf{H_Y}$ are fed into the Query and Basic block stacks, followed by a 2D convolution layer with $3 \times 3$ kernel and a ReLu activation to output the target speech mask element multiplied with $\mathbf{H_Y}$. Finally, a 2D convolution layer with $3 \times 3$ kernel is used to map the TF embeddings into real and imaginary components, and the obtained complex spectrogram is converted back into time-domain using the inverse STFT. This section introduces these modules.
\subsubsection{Query block}
The Query block contains two Query embedding generators (QEG) for learnable query embeddings $\mathbf{E_{Q1}}$ and $\mathbf{E_{Q2}}$, learned from query distance and room features, as shown in Fig~\ref{fig:two} (b). These embeddings are fused with the intermediate representation in both temporal and spectral domains, which can be regarded as retrieving the learned query embeddings in all TF bins among intra-subband and intra-frame directions. In detail, for the intra-subband process, the input TF embedding $\mathbf{H_Y}$ is first concatenated with the query embedding $\mathbf{E_{Q1}}$ along the temporal axis. This combined embedding is passed through Layer Normalization (LN) along its last dimension and subsequently processed by a Bidirectional Long Short-Term Memory (BLSTM) layer. Finally, the last concatenated dimension in the hidden state of the BLSTM is cropped to match the input dimension and then passed through a linear layer that is activated using a Gaussian error linear unit (GELU) and connected using residual. The intra-frame fusion follows the same procedure, but concatenates the query embedding $\mathbf{E_{Q2}}$ along the spectral dimension, and applies the BLSTM across subbands in each frame.
\subsubsection{Query embedding generator}
The QEG converts the raw query distance and room information into useful query embeddings. The structure of QEG is presented in Fig~\ref{fig:two} (c). The used clues are raw query distance $d_q$, and room clues including microphone-wall distances sequence $Dis_{mw}=[dis_1, dis_2, ..., dis_6]$, and reverberation time $RT60$, each of which corresponds to one linear layer and outputs three embeddings $E_{dis}$, $E_{dim}$, and $E_{rt}$, these embeddings are concatenated and fed into three linear layers with tanh activation function for final query embedding $\mathbf{E_Q}$ with $D$-dimension \cite{Gu-ReZeroRegioncustomizableSoundExtraction-2023}. We use the distance from the microphone to surrounding walls as features because it represents the room dimensions and microphone placement simultaneously, and it has been shown that the sound field transfer function can reflect the structure of the surrounding space, e.g., the distance to the nearest wall \cite{luo2022learning}. The $E_{dis}$ embedding is calculated by summing all individual mic-wall distance embeddings, and this ambiguous representation shows better generalization performance. 
\subsubsection{Basic block}
The Basic block adopts the same structure as the Query block but omits the fusion step, applying a single RNN across all frames and subbands, similar to DPRNN \cite{Luo-DualpathRNNEfficientLong-2020}.  
\subsection{Loss function}
  Two kinds of loss functions are used for the model to output both target speech and zero speech conditioned on the query distance. Given the target speech $\mathbf{x}_{s}$ and the estimated speech $\hat{\mathbf{x}_s}$, we adopt the signal-to-distortion ratio (SDR) as the loss function when the speakers exist at the query distance,
\begin{align}
    L_{active}(\mathbf{x}_{s}, \hat{\mathbf{x}_s}) = 10 \log_{10} \left( \frac{\|\mathbf{x}_s\|^2}{\| \mathbf{x}_s - \hat{\mathbf{x}_s} \|^2 + \tau \|\mathbf{x}_s\|^2} \right)
\end{align}
  \noindent
  where the $\tau$ is the soft threshold to control the upper bound of the loss and set to $10^{-3}$. The scale invariant SDR (SI-SDR) \cite{Roux-SDRHalfbakedWellDone-2019} is not considered since we wish the model could handle inactive source conditions in which the model would output zero values \cite{Delcroix-SoundBeamTargetSoundExtraction-2022}. For the condition that no speaker is near the query distance, the adopted loss function is $L_{0}$
  \cite{Wisdom-WhatAllFussFree-2021, Delcroix-SoundBeamTargetSoundExtraction-2022},
  \begin{align}
    L_{0}(\mathbf{y}, \hat{\mathbf{x}_s}) = 10 \log_{10}\left({\| \hat{\mathbf{x}_s}\|^2 + \tau^\text{inactive} \| \mathbf{y} \|^2}\right),
  \end{align}
  \noindent
  where $\mathbf{y}$ is the mix speech, and the $\tau^{\text{inactive}}$ represents
  the soft threshold which is set to $10^{-2}$.
  \begin{figure}[!t]
    \begin{minipage}[b]{1.0\linewidth}
      \centering
      \centerline{\includegraphics[width=8.0cm]{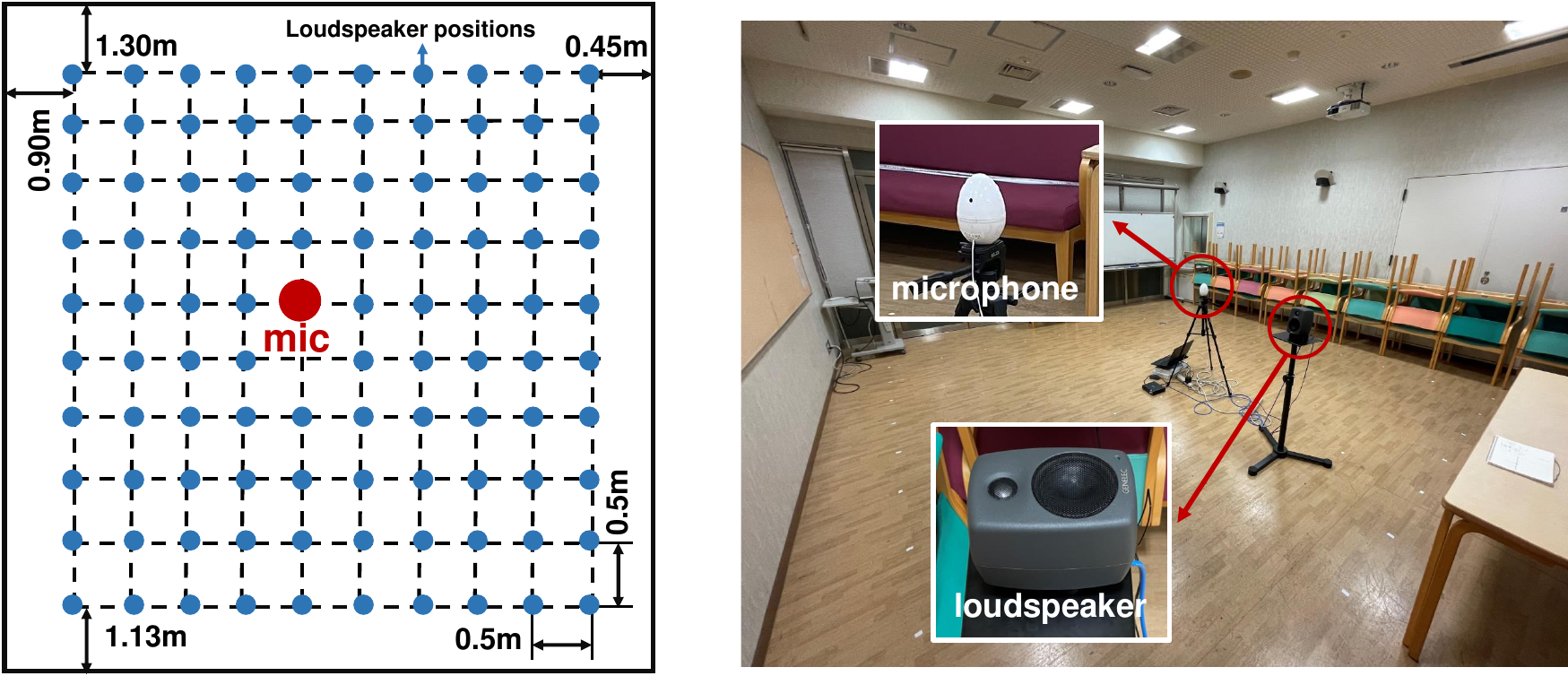}}
    \end{minipage}
    \caption{RIR recording locations and environment.}
    \label{fig:rir}
    \vspace{-1em}
  \end{figure}
  \section{Experiment}
  \label{sec:pagestyle}
  \subsection{Dataset Generation}
For model training, the mixed speech is formed by superposing multiple reverberant signals, each generated by convolving an anechoic speech signal with a location-specific RIR, and we consider both simulated and realistic RIRs. Additionally, real-world speech is also recorded for further validation.
  \\
   \textbf{Simulated RIR dataset}: We used the randomized image method for RIR generation \cite{Scheibler-PyroomacousticsPythonPackageAudio-2018}.
  Specifically, considering different room settings and microphone placement
  significantly influenced the transfer function characteristics and the model performance, we constructed two sub-simulated RIR datasets to assess the model's generalization capabilities.
  \subsubsection{\textbf{Sim1} One room with fixed microphone position}
  This dataset contains a $7 \times 8 \times 3$ meters room with a fixed microphone at $3.5 \times 4 \times 1.1$ meters. The reverberation time (RT60) is set to 0.2s. Speaker positions are randomly initialized at least 0.5 meters from the walls, with heights ranging from 1.2 to 2.0 meters. 10,000 RIRs were generated within 0.2 to 5.0 meters from the microphone.
  \subsubsection{\textbf{Sim2} 1,000 rooms}
  This simulated dataset contains 1,000 randomly generalized rooms between sizes $4 \times 5 \times 2.5$ meters to $8 \times 10 \times 3.0$ meters. The RT60 is randomly selected between 0.2s to 0.5s. Each room has one microphone position, each corresponding to 500 speaker positions, resulting in 490,325 RIR samples. To ensure a more balanced distribution of samples across varying distances, the generation process is organized into distance bands, e.g., 0-0.5 meters, so the RIRs for farther speakers may be missing in smaller rooms.
  \\
  \textbf{Real recorded dataset}: Open source datasets of location-dependent RIRs or real recordings are scarce. To make the evaluation more realistic, we collected real RIRs and speech samples in a conference room with a reverberation time of 0.6 seconds. The room size is $5.9 \times 6.9 \times 2.9$ meters. 
    \setcounter{subsubsection}{0}
    \subsubsection{\textbf{Real recorded RIR}}
 We collect real RIRs in a meeting room using the sine sweep method and follow the processing from \cite{Farina-AdvancementsImpulseResponseMeasurements-2007}. We use an eight-channel microphone array to record the RIR at 98 locations, as shown in Fig~\ref{fig:rir}, resulting in 784 single-channel RIR segments. 
  This dataset is publicly released for community research. 
    \footnote{\url{https://github.com/RunwuShi/GridRIRDataset}}
    \subsubsection{\textbf{Real-world recorded speech}}
We collect real speech signals in the same room. Specifically, we fix the microphone and play speech signals from the LibriSpeech dataset at 14 distinct locations with different distances. For each location, we play ten speech segments from the same speaker. Since the speech is recorded separately, supervised training can be realized by overlapping different recorded speech.
\\
   \textbf{Speech and sample generation}: For the Sim1, Sim2, and real recorded RIR datasets, the speech is adopted from the LibriLight \cite{Kahn-LibriLightBenchmarkASRLimited-2019} dataset, and 128, 48, and 64 speakers were used for training, validation, and testing. Simulated RIR datasets were split into training, validation, and testing datasets with a ratio of 0.9:0.02:0.08. Each utterance with a length of 4s is convolved from randomly chosen speech and RIR, and the root mean square energy is randomly adjusted from -25 to -20 dB to simulate the louder voices of farther speakers.
  \subsection{Training Process and Configuration}
  The model contains 4 Query blocks and 4 Basic blocks, the dimension size $D$ 
 and the hidden state size of LSTM is set to 64. The three linear layers in each QEG contain 96, 64, and 64 units. A frame length of 32 ms with 16 ms frameshift is used in STFT. The $r_{spk}$ centered with ground truth speaker distance is set to 0.5 meters. For training, we use the Adam \cite{Kingma-AdamMethodStochasticOptimization-2017} optimizer, and gradient clipping is set to a maximum norm of 5. The batch size is set to 14. The initial learning rate is 0.001, and if no lower loss is found in 10 epochs, the learning rate is reduced to 80\%. Training has 400 epochs with 25\% inactive samples corresponding to zero output.
  \setlength{\textfloatsep}{5pt plus 1.0pt minus 1.0pt}
\begin{table}[!t]
    \caption{Results of two speaker mixtures on Sim1 and Sim2. SDR and SDRi report \textbf{non-overlap}/overlap speaker conditions.}
    \label{tab:one}
    \centering
    \renewcommand{\arraystretch}{1.3} 
    \begin{threeparttable}
      \begin{tabularx}
        {\columnwidth}
        { >{\arraybackslash}p{0.6cm} >
        {\centering\arraybackslash}p{1.25cm} >{\centering\arraybackslash}p{1.2cm} >{\centering\arraybackslash}p{1.2cm} >{\centering\arraybackslash}p{0.5cm} >{\centering\arraybackslash}p{0.5cm} >
        {\centering\arraybackslash}p{0.8cm} }
        \specialrule{1pt}{1pt}{1pt} 
        Training Dataset & Clue & SDR$\uparrow$ (dB) & SDRi$\uparrow$
        (dB) & PESQ$\uparrow$ & $L_{0} \downarrow$ & No/o ratio\\ 
        \midrule 
        \textbf{Sim1} & Dis & $\scalebox{1.3}
        {\nicefrac{11.95\,}{\,\textnormal{24.48}}}$ &
        $\scalebox{1.3}
        {\nicefrac{11.99\,}{\,\textnormal{-5.52}}}$ &
        2.98 & 4.56 & 20.40\% \\
        \midrule 
        \multirow{4}{*}[-0mm]{\centering \textbf{Sim2}} 
        & Dis & $\scalebox{1.3}{\nicefrac{6.71\,}{\,\textnormal{18.07}}}$ &
        $\scalebox{1.3}{\nicefrac{6.62\,}{\,\textnormal{-11.93}}}$ &
        2.34 & 20.35 & 20.30\%\\
        & \textbf{Dis+Dim+Rt }
        & $\scalebox{1.3}{\nicefrac{{7.96}\,}{\,\textnormal{19.03}}}$ 
        &$\scalebox{1.3}{\nicefrac{{8.08}\,}{\,\textnormal{-10.97}}}$ 
        & 2.54 & 12.52 & 20.20\%\\ 
        & Dis+Rt 
        & $\scalebox{1.3}
        {\nicefrac{7.46\,}{\,\textnormal{19.35}}}$ 
        & $\scalebox{1.3}
        {\nicefrac{7.45\,}{\,\textnormal{-10.65}}}$ &
        2.46 & 16.04 & 20.08\%\\ 
        & Dis+Dim 
        & $\scalebox{1.3}{\nicefrac{7.13\,}{\,\textnormal{19.99}}}$ 
        & $\scalebox{1.3}{\nicefrac{7.17\,}{\,\textnormal{-10.01}}}$ &
        2.42 & 16.09 & 20.44\%\\
        \specialrule{1.2pt}{1pt}{1pt}
      \end{tabularx}
    \end{threeparttable}
\end{table}
\begin{table}[!t]
    \caption{Generalization performance on real collected RIR dataset.}
    \label{tab:two}
    \centering
    \renewcommand{\arraystretch}{1.3} 
    \begin{threeparttable}
      \begin{tabularx}
        {\columnwidth}
        {>
        {\arraybackslash}p{0.6cm} >
        {\centering\arraybackslash}p{1.25cm} >{\centering\arraybackslash}p{1.2cm} >{\centering\arraybackslash}p{1.2cm} >{\centering\arraybackslash}p{0.5cm} >{\centering\arraybackslash}p{0.5cm} >
        {\centering\arraybackslash}p{0.8cm} }
        \specialrule{1.2pt}{1pt}{1pt} 
        Training Dataset & Clue & SDR$\uparrow$ (dB) & SDRi$\uparrow$ (dB) & PESQ$\uparrow$ & $L_{0} \downarrow$ & No/o ratio\\
        \midrule 
        \multirow{1}{*}[0mm]{\textbf{Sim1}}
        & Dis & $\scalebox{1.3}
        {\nicefrac{0.71\,}{\,\textnormal{5.32}}}$ &
        $\scalebox{1.3}
        {\nicefrac{0.74\,}{\,\textnormal{-24.68}}}$ &
        1.56 & 14.29 & 38.56\%\\ 
        \midrule 
        \multirow{4}{*}[0mm]{\textbf{Sim2}}
        & Dis & $\scalebox{1.3}
        {\nicefrac{3.06\,}{\,\textnormal{10.32}}}$ &
        $\scalebox{1.3}
        {\nicefrac{3.00\,}{\,\textnormal{-19.68}}}$ &
        1.95 & 21.37 & 40.64\%\\ 
        & \textbf{Dis+Dim+Rt} & $\scalebox{1.3}
        {\nicefrac{{4.50}\,}{\,\textnormal{9.46}}}$ &
        $\scalebox{1.3}
        {\nicefrac{{4.42}\,}{\,\textnormal{-20.54}}}$ &
        2.11 & 14.30 & 39.06\%\\ 
        & Dis+Rt & $\scalebox{1.3}
        {\nicefrac{4.01\,}{\,\textnormal{10.40}}}$ &
        $\scalebox{1.3}
        {\nicefrac{4.02\,}{\,\textnormal{-19.60}}}$ &
        2.04 & 19.52 & 40.54\%\\ 
        & Dis+Dim & $\scalebox{1.3}
        {\nicefrac{3.35\,}{\,\textnormal{7.32}}}$ &
        $\scalebox{1.3}
        {\nicefrac{3.40\,}{\,\textnormal{-22.68}}}$ &
        1.94 & 16.13 & 38.62\%\\ 
        \midrule 
        \textbf{Finetune} & Dis & $\scalebox{1.3}{\nicefrac{8.20\,}{\,\textnormal{11.42}}}$ &
        $\scalebox{1.3}{\nicefrac{8.06\,}{\,\textnormal{-18.58}}}$ &
        2.57 & 14.43 & 8.84\%\\ 
        \specialrule{1.2pt}{1pt}{1pt}
      \end{tabularx}
    \end{threeparttable}
\end{table}
\section{Results and Discussion}
\subsection{Results on Simulated RIR Dataset}
The evaluation metric needs to consider both the presence and absence of speakers at the query distance. For the presence condition, we use SDR, SDR improvement (SDRi), and Perceptual Evaluation of Speech Quality (PESQ) to verify speech quality and scale accuracy. For the absence condition, we use the inactive loss $L_{0}$ as the metric. Each test generates 1,000 utterances randomly, and the results of each dataset are obtained by testing five times and expressed as the mean. Specifically, for the presence condition of existing multiple speakers near the query distance, the model will output the overlapped speech, and we report the SDR and SDRi under the non-overlapped and overlapped speaker conditions separately. We also present the ratio of non-overlap/overlap samples.

Table~\ref{tab:one} presents the results of distance-based TSE under conditions with and without adding room information as the clues, and we test both models trained on two speaker mixtures from the Sim1 and Sim2 datasets. Sim1 scenario can be regarded as the most idealized experiment setting, in which the model is trained and tested in the same room with a fixed microphone position setup, and this performance should be the best performance of such distance-based TSE methods. For the Sim1 dataset containing one room, the TSE using only distance clue achieves an SDR of 11.95 dB under non-overlapped speaker conditions. For the Sim2 dataset containing 1,000 rooms, we evaluate the performance of distance-based TSE methods using various additional room-related clues, which utilize room configuration (Dis+Dim), reverberation time (Dis+Rt), and a combination of both as auxiliary clues (Dis+Dim+Rt). The result presents the model using room configuration and reverberation time achieves an SDR of up to 7.96 dB. Moreover, lower $L_0$ indicates that the use of room clues effectively improves recognition of non-existent speakers, which is essential for speaker distance estimation \cite{Kushwaha-SoundSourceDistanceEstimation-2023, shi2024distance, Neri-SpeakerDistanceEstimationEnclosures-2024}. The results also show that the reverberation information improves the quality of the extracted speech more than the microphone-wall distances. Overall, the different rooms introduced by Sim2 have a great performance decay relative to the one room of Sim1, suggesting the necessity of additional room information.
\subsection{Performance on Real Recorded RIR and Speech Dataset}

This section first evaluates the generalization ability of several distance-based TSEs on the real collected RIR datasets. The results are presented in Table~\ref{tab:two}. For distance clue-only methods, the Sim2-trained model achieves an SDR of 3.06 dB, outperforming the model trained on Sim1, which contains one room. For the methods with additional room clues, the SDR can be further improved to 4.50 dB. We also finetune the model trained on Sim2 by 100 epochs using 70\% of the real RIR dataset and set $r_{spk}$ to 0.1m for more accurate extraction. The SDR can be further enhanced to 8.20 dB. These results demonstrate the effectiveness of the proposed method in real RIR scenarios.
\begin{table}[!t]
    \caption{Performance of finetuned model on real recorded speech}
    \label{tab:three}
    \centering
    \renewcommand{\arraystretch}{1.3} 
    \raggedleft
    \begin{threeparttable}
      \begin{tabularx}
        {\columnwidth}
        { >{\arraybackslash}p{1.5cm} >
        {\centering\arraybackslash}p{0.9cm} >{\centering\arraybackslash}p{0.9cm} >{\centering\arraybackslash}p{1.1cm} >{\centering\arraybackslash}p{1.1cm} >
        {\centering\arraybackslash}p{1.0cm} }
        \specialrule{1pt}{1pt}{1pt} 
         Finetune Type & SDR $\uparrow$ (dB) & SDRi $\uparrow$
        (dB) & SI-SDR $\uparrow$ (dB) & SI-SDRi $\uparrow$ (dB) & PESQ$\uparrow$\\ 
        \midrule  
        All location & 7.24 & 7.22 & 5.19 & 5.17 & 2.46 \\ 
        New location & 7.34 & 7.31 & 5.57 & 5.55 & 2.45\\         
        \specialrule{1.2pt}{1pt}{1pt}
      \end{tabularx}
    \end{threeparttable}
\end{table}

To verify the effectiveness of the distance-based TSE methods on real-world speech, we also finetune the model on real recorded reverberant speech segments. Finetuning is necessary because both simulated and realistic RIRs differ from actual reverberant conditions, especially given the sensitivity of distance cues to environmental factors.  Specifically, we compare two strategies, in the first strategy, we use all of the locations to finetune the model, where 50\% of the audio is used with finetune and the other 50\% is used for testing, and in the second strategy, we use all of the audio from the 7 locations for finetune and use the audio from the others locations for testing. The finetune process contains 10 epochs with a learning rate of 0.0001. The results are presented in Table~\ref{tab:three}. The distance-based TSE can achieve an SDR of 7.34 dB under the second fintune strategy. 

\begin{table}[!t]
    \caption{Comparison of Distance and Enrolled voice based TSE on Sim2}
    \label{tab:four}
    \centering
    \renewcommand{\arraystretch}{1.3} 
    \begin{threeparttable}
      \begin{tabularx}
        {\columnwidth}
        { >{\arraybackslash}p{1.2cm} >
        {\centering\arraybackslash}p{0.65cm} >{\centering\arraybackslash}p{0.8cm} >{\centering\arraybackslash}p{0.8cm} >{\centering\arraybackslash}p{0.9cm} >{\centering\arraybackslash}p{1.0cm} >
        {\centering\arraybackslash}p{1.0cm} }
        \specialrule{1pt}{1pt}{1pt} 
         Method & Params ($\times10^{6}$) & SDR $\uparrow$ (dB) & SDRi $\uparrow$
        (dB) & SI-SDR $\uparrow$ (dB) & SI-SDRi $\uparrow$ (dB) & PESQ$\uparrow$\\ 
        \midrule  

        Dis & 1.25 & 6.71 & 6.62 & 5.46 & 5.37 & 2.34 \\ 
        Dis+Dim+Rt & 1.29 
        & 7.96 & 8.08 & 6.94 & 7.05 & 2.54\\         
        \midrule 
        SpEx+~\cite{Ge-SpExCompleteTimeDomain-2020} & 11.15
        & -- & -- & 3.59 & 3.39 & 1.85
        \\
        CIENet~\cite{Yang-TargetSpeakerExtractionDirectly-2024b} & 2.67 & -- & -- &
        9.19 & 8.90 & 2.72\\ 
        \specialrule{1.2pt}{1pt}{1pt}
      \end{tabularx}
    \end{threeparttable}
\end{table}
\subsection{Performance Comparison with Different Methods}
As shown in Table~\ref{tab:four}, the proposed method is compared with other methods on the Sim2 dataset. We reproduce two advanced enrolled speech-based TSE models, a time-domain TSE model SpEx+ \cite{Ge-SpExCompleteTimeDomain-2020} and a TF-domain TSE model CIENet-mDPRNN \cite{Yang-TargetSpeakerExtractionDirectly-2024b}, for which we used 8 seconds of target speaker voice as the clue for both models and the training objective and evaluation metrics follow the original papers. We adopt the official implementation of SpEx+ and implement CIENet ourselves. For a fair comparison, we test the distance-based methods where each query distance corresponds to only one speaker. The results demonstrate the feasibility of distance-based TSE. However, due to the heterogeneity between the distance clue and speech, extracting speech is inherently more challenging than using the enrolled voice. 

  \section{Conclusion}
  We proposed to utilize distance and room information as clues for the TSE system, and we built realistic datasets to verify the proposed method. The results demonstrate the use of such room information enhances the performance and generalization ability of distance-based TSEs, and the performance on real recorded speech also proves the potential of this task.  Future work will focus on more real-world scenarios.
  \bibliographystyle{IEEEtran}
  \bibliography{ref}
\end{document}